# Acceleration of diffusion in ethylammonium nitrate ionic liquid confined between parallel glass plates

Andrei Filippov,[a,b,*] Oleg I. Gnezdilov,[b] Nicklas Hjalmarsson,[c] Oleg N. Antzutkin,[a] Sergei Glavatskih,[d,e] Istvan Furó,[f] and Mark W. Rutland[c,g]

[a]*Chemistry of Interfaces, Luleå University of Technology, SE-97187 Luleå, Sweden*
[b]*Institute of Physics, Kazan Federal University, 420008 Kazan, Russia*
[c]*Surface and Corrosion Science, KTH Royal Institute of Technology, SE-10044 Stockholm, Sweden*
[d]*System and Component Design, KTH Royal Institute of Technology, SE-10044 Stockholm, Sweden*
[e]*Department of Mechanical Construction and Production, Ghent University,*

*B-9000 Ghent, Belgium*
[f]*Applied Physical Chemistry, KTH Royal Institute of Technology, SE-10044 Stockholm, Sweden*
[g]*SP Technical Research Institute of Sweden, Chemistry Materials and Surfaces, Box 5607, SE-114 86 Stockholm, Sweden*

*E-mail: Andrei.Filippov@ltu.se

The bulk self-diffusion of the ethylammonium cation (measured by $^1$H NMR) and the nitrate anion (measured by $^{15}$N NMR) can both be described by respective single diffusion coefficients, of which that of the anion is 1.7-times higher than that of the cation. This indicates no tight association of the ions in the ionic liquid. For the ethylammonium cation (EA) of the EAN confined between glass plates the effective diffusion coefficient along, as well as normal to the confining glass plates is higher by a factor of 1.86 as compared to that in the bulk. The same time, $^1$H NMR $T_2$ relaxation of protons of –NH$_3$ group of the EA cation is faster by a factor of ~ 22 than that in bulk. $^2$H NMR spectra of selectively labeled –CH$_2$-and -CH$_3$ groups of EA do not demonstrate any ordering of the EA between the glass plates. We suggested that these data favor a model where a bulk isotropic sponge-like structure of EAN is saved in the confinement, but sizes of connecting channels increases. Those leads to faster translational diffusion and faster exchange processes of protons of –NH$_3$ group, in comparison with the bulk.

Ionic liquids (ILs) are molten salts formed typically of organic cations and either organic or inorganic anions.[1,2] Their applications are continuously expanding, for example as electrolyte material in lithium batteries[3] and ultracapacitors,[4] as media for chemical reactions and separation,[2,5] and as lubricants.[6] Ethylammonium nitrate (EAN), first synthesized by Paul Walden in 1914,[7] is the most commonly reported protic IL.[1] It is used as a replacement for organic solvents as a reaction medium, as a precipitating agent for protein crystallization,[5] an electrically conductive solvent in electrochemistry,[3] amongst other applications. Similarly to water, EAN has a three-dimensional hydrogen-bonding network and can be used as an amphiphilic self-assembly medium.[8] Bulk organization of EAN is often characterizes as bi-continuous, sponge-like structure.[9-11]

Recently, small-angle neutron scattering revealed that EAN itself exhibits an inherent amphiphilic nanostructure in the pure liquid state.[9] For the lamellar structures deemed as most probable, the calculated Bragg spacing is approximately twice the ion-pair dimension of the IL, which suggests that the IL is structured on the length scale of the ions, with the (hydrogenous) alkyl groups associated together and segregated from the H-bonded ionic moieties $-NH_3^+$ (or $-ND_3^+$) and $NO_3^-$. Based on this finding, it has been suggested that surfaces may induce alignment of such locally ordered domains, creating a persistent molecular layering.[10] Vibrational sum frequency spectroscopy and X-ray reflectivity studies[11] confirmed the existence in EAN of significant interfacial structures within a *ca* 3 nm thick layer. Surface structures have also been found in other ILs,[12-14] sometimes persisting up to tens of nanometers into the bulk[13,14] with a clear dependence on the molecular nature of the surface. Exceptionally extensive surface-induced structures with a thickness up to 2 µm were recently demonstrated for a number of imidazolium bis(trifluoromethylsulfonyl)imide ILs by Anaredy and Shaw.[15] Much less attention has been paid to the structure and dynamics of ILs, which can be formed in a micrometer-scale confinement.

The purpose of this work was to study the dynamic features exhibited by EAN in the presence of polar surfaces and micrometer-scale confinement. For this purpose, we exploit NMR diffusion measurements, previously applied to a wide range of ILs.[16-22] Of particular relevance, diffusion NMR has been used to study the molecular dynamics in ILs under confinement (coincidentally, all in silica pores).[18,23-27] In comparison to NMR relaxation studies, diffusion NMR data are typically simpler to interpret and are more robust regarding experimental artifacts,[28,29] both of which are of great advantage considering the complex behavior ILs are rather prone to show. In contrast to other transport methods, NMR diffusion data can be obtained selectively for the different constituting moieties, in the present case the nitrate anion and the ethylammonium cation. This latter feature, as also shown here, can be aided by suitable isotope enrichments ($^{15}N$ or/and $^2H$).

EAN, shown in Fig.1, consists of an ethylammonium (EA) cation and a nitrate (NI) anion. All samples were synthesized following previously established recipes.[30,31] While Sample 1 of EAN has a natural abundance (n.a.) of all isotopes: $^1H$ (99.98%), $^{13}C$ (1.108%) and $^{15}N$ (0.37 atom %) in both EA and NI, the nitrate anion was $^{15}N$ enriched (~ 98 atom %) in Sample 2 of EAN. Sample 3 of EAN was prepared by admixing EAN-$D_5$, i. e. selectively deuterated at $CD_3$- and $CD_2$- positions of the cation, into Sample 1 at a composition 20/80 mol% of EAN-$D_5$/EAN(n.a.). Prior to experiments, each sample was degassed under vacuum (pressure less than $10^{-3}$ mbar, temperature 313 K) for 60 hours. The chemical composition of each sample (including impurities) was established by liquid $^1H$ and $^{13}C$ NMR. In summary, the hydrogen and carbon content associated with the ethylammonium cation were approximately 99% in Sample 1 and larger than 99.6% in Sample 2. In a series of experiments, samples of EAN were contained between thin glass plates arranged in stacks (Fig.2).

The plates used (5 x 14 x 0.1 mm, Thermo Scientific Menzel Gläser, Menzel GmbH, Germany) were cleaned carefully. Contact angle measurements with Milli-Q water provided a contact angle near 0°, indicating hydrophilic glass surfaces. A drop of EAN was placed on the first plate, then it was covered by the second plate with placing another drop of EAN on the top, and so on until the thickness of the stack reached approximately 5 mm. Excess of EAN from sides of the stack was removed by wiping. The plates were thereafter placed in a sample cell of square cross section. The mean distance (spacing) between glass plates was estimated by weighing the introduced EAN, which yielded *d* ~3.8-4.5 µm. Measuring the thickness directly and subtracting from it the total glass thickness indicates that this is a consistent value.

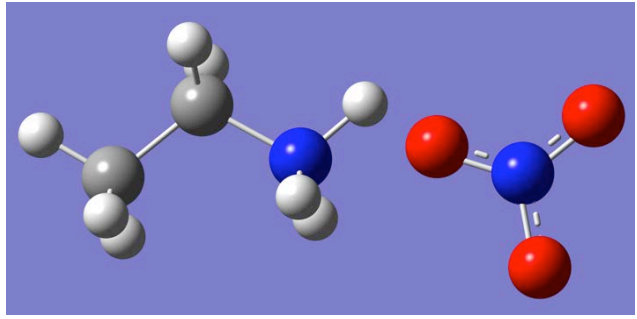

**Figure 1.** The chemical structure of EAN consisting of an ethylammonium cation and a nitrate anion, with nitrogen (blue), oxygen (red), carbon (dark grey) and hydrogen (light grey) atoms.

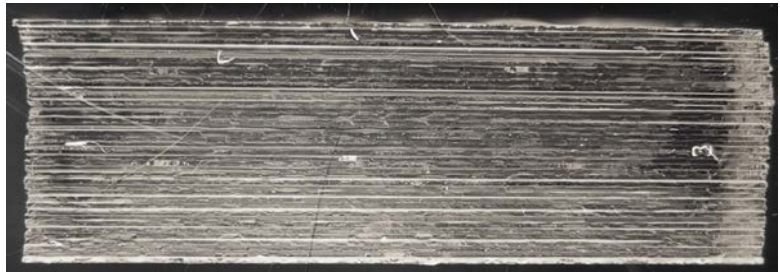

**Figure 2.** Alignment of glass plates with confined EAN.

$^1$H (Samples 1 and 2) and $^{15}$N (Sample 2) NMR self-diffusion measurements of EAN in bulk were performed with a Pulsed-Field-Gradient (PFG) probe Diff50 (Bruker) and a Bruker Ascend Aeon WB 400 (Bruker BioSpin AG, Fällanden, Switzerland) NMR spectrometer was used with working frequency 400.27 MHz for $^1$H and 40.56 MHz for $^{15}$N. Data were processed using Bruker Topspin 3.5 software. $^2$H NMR spectra of Sample 3 were obtained with the same spectrometer using the quadrupole-echo pulse sequence. $^1$H $T_1$ and $T_2$ NMR relaxation time measurements were performed with inversion-recovery (180°-τ-90°-fid) and CPMG (90°-τ-180°-τ-echo) pulse sequences, and analyzed using Bruker Topspin 3.5 software.

$^1$H NMR diffusion measurements of EAN confined in the glass stack were performed with a NMR solenoid insert, which was used to macroscopically align the films in the sample stack at angles 0° and 90° with respect to the applied magnetic field gradient, the latter being co-parallel to the main magnetic field.

The diffusional decays (DD) were recorded using the stimulated echo pulse sequence. For single-component isotropic diffusion, the signal intensity $A$ changes as:[32,33]

$$A(\tau,\tau_1,g,\delta) \propto \exp\left(-\frac{2\tau}{T_2} - \frac{\tau_1}{T_1}\right) \exp\left(-\gamma^2 \delta^2 g^2 D t_d\right) \quad (1)$$

where $T_1$ and $T_2$ are the spin-lattice and the spin-spin relaxation times, respectively; $\tau$ and $\tau_1$ are time intervals in the pulse sequence; $\gamma$ is the gyromagnetic ratio for a used nucleus; $g$ and $\delta$ are the amplitude and the duration of the gradient pulse; $t_d = (\Delta - \delta/3)$ is the diffusion time;

$\Delta=(\tau + \tau_1)$; and $D$ is the self-diffusion coefficient. In the experiments the gradient strength $g$ was varied with all other parameters kept constant. If not stated otherwise, the $D$ values were obtained by fitting Eq. 1 to the experimental decays.

The $^1$H and $^{13}$C NMR spectra of EAN arise exclusively from and report about the EA cation since the anion lacks both protons and carbon. These spectra demonstrated three resonance lines, which were assigned to the $NH_3^+$, $CH_2$ and $CH_3$ moieties, in accordance with previously published data.[34] As expected, the $^{15}$N NMR spectrum of the nitrate anion in Sample 2 revealed a single broad resonance line at *ca* 380 ppm, in accordance with literature data.[35]

The DDs of the $^1$H signals of the EA cation in Sample 1 and in Sample 2, as well as the DD of the $^{15}$N signal of the nitrate anion, are all linear in the semi-logarithmic scale. Therefore, the mobility of each ion can be described by a single diffusion coefficient. An important observation is that the $D$ values of the nitrate anion are by a factor of ~1.7 higher than those of the EA cation. This means that the diffusive motion of cations and anions in bulk EAN is not tightly coupled (that is, ion pairing is not dominant). In dense ionic systems, self-diffusion is influenced strongly by electrostatic interactions and is no longer defined solely, as in simple aprotic liquids built up by neutral molecules, by molecular size.[21] In EAN, hydrogen bonding, too, is a significant intermolecular force with concomitant solvophobicity.[9] To exemplify this point, one should recall that the viscosity of EAN is *ca* 30 times higher than that of e.g. nitropropane or similar liquids.[36,37] In addition, EAN exhibits a nano-scale order[9] instead of a random local structure. Hence, it is difficult to pinpoint a single dominant reason for the factor-1.7 difference. We note that in another IL with nano-scale ordering one found in magnitude rather similar (factor-2) ratio between anion and cation diffusion coefficients.[22] Yet, in that system, this ratio seemed to depend strongly on the nano-scale order that clearly shows the importance of local structure for translational dynamics of the ions.

Fig. 3 shows the DDs of the $^1$H signals of the EA cation with the field gradient directed along and normal to glass plates. Under current conditions, the DDs recorded are caused by diffusive molecular displacements in the IL films along and normal to the glass plates, respectively. In the direction *along* the plates (Fig.3A) the DDs maintain the linear form typical of the bulk state, but the slope (dotted line) of the decays and, consequently, the value of the derived self-diffusion coefficient $D_\parallel = D^*$ is a factor of 1.86 higher than $D_0$ in the bulk (the decay for which is also shown for comparison). The DD is invariant to changing the diffusion time in the range of 50 - 1000 ms.

In the direction *normal* to the plates (Fig.3B, colored symbols) and short diffusion time (3 ms and 10 ms) the signal decays faster than that in bulk (black squares), while it approaches to the slope corresponding to $D^*$, observed for diffusion along the plates. This means that diffusion is accelerated by similar extent along as well as normal to the plates, $D_\perp = D_\parallel = D^*$.

Forms of DDs (Fig.3B) deviate from the single-exponential behavior typical for the bulk (recall Eq.(1), shown here as a solid line and black squares) and the extent of this deviation increases as the diffusion time increases from 3 ms to 1 s. Apparent slopes of DDs normal to the plates at diffusion times higher 10 ms are less than those in the bulk and get lesser by increasing diffusion time, it is clear that molecular displacements across the IL films are hindered.[28,32] The hindrance is constituted by the glass plates whose separation of *ca* 4 µm is comparable to the diffusional path-length of $L \approx \sqrt{(D_0 \cdot t_d)}$,[38] which is, in the current case, between 1.7 and 7.5 µm.

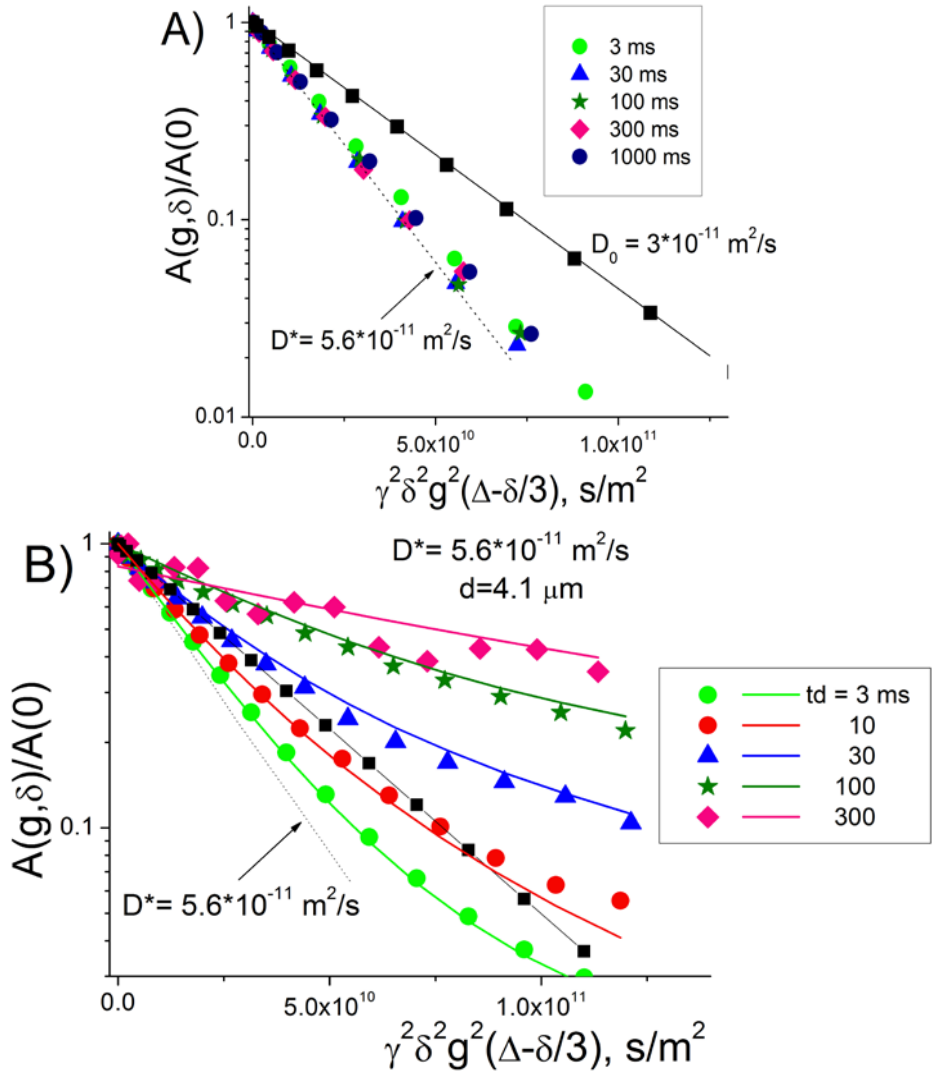

**Figure 3.** The DDs of the $^1$H NMR signals of the EA cation recorded at 293 K by pulsed-field-gradient stimulated echo experiments in bulk EAN (solid squares, $D_0 = 3.0 \cdot 10^{-11}$ m$^2$/s) and in EAN films confined between parallel glass plates, with gradient directions along (**A**) and perpendicular to (**B**) the plates. The DDs were obtained by increasing the gradient amplitude **g** and the different colored symbols indicate data recorded at different diffusion times $t_d$. Best fits (Eq. (2)) of suitable expressions for restricted diffusion to the data in (**B**) are presented by lines of corresponding colors. Dashed line in (**B**) corresponds to $D^* = 5.6 \cdot 10^{-11}$ m$^2$/s.

As is well known, in such situations the mean displacement levels off, while the apparent mean diffusion coefficient decreases with increasing $t_d$ revealing the same qualitative behavior as shown in Fig. 3B. Planar restriction (diffusion along plane normal between parallel plates) is one of simplest regular geometries for which the DD of confined liquids has been analytically solved,[39] under the assumption of elastic collisions of diffusing molecules with the wall. The expression for the DD can be presented in the following form, which exactly describes diffusion decays of molecular liquids for all regimes of diffusion in this geometry (see Eq.(S2)):[39]

$$A(\delta, \Delta, g) = \frac{2[1-\cos(\gamma g \delta d)]}{(\gamma g \delta d)^2} + 4(\gamma g \delta d)^2 \times \sum_{n=1}^{\infty}\left\{\frac{1-(-1)^n \cos(\gamma g \delta d)}{[(n\pi)^2 - (\gamma g \delta d)^2]^2} \times \exp\left(-\frac{n^2 \pi^2 D^* \Delta}{d^2}\right)\right\},$$ (2)

where $D^*$ is the diffusion coefficient "un-distorted" by collisions with walls. Scrutiny of both the form and diffusion time dependences of the experimental DDs for diffusion normal to barriers (see Fig.3B) reveals that there is a dependence of the DDs on $t_d$, which is typical of the intermediate diffusion time regime. $L$ calculated as $L \approx \sqrt{(D_0 \cdot t_d)}$ is in the range 1.7 - 7.5 µm and is comparable with the plate spacing, ~ 4 µm. This also corresponds to the intermediate diffusion time regime. Therefore, an iterative procedure (Eq. (2)) was applied. The equation was solved with the number of iterations being varied up to 1000. The separation between planes was first estimated by weighing and thickness measurements as mentioned above, but then used as a fitting parameter, together with $D^* = 5.6 \cdot 10^{-11}$ m$^2$/s, to better match the experimental time-dependent DDs. One of the peculiarities of diffusion is that for very regular distances between planes, DDs usually demonstrate a so called "diffusion diffraction" effect, i.e. periodic oscillations on DDs.[39] No such oscillations on DDs were detected in these experiments (Fig.3B). According to a previous study,[32] "diffusion diffraction" effects should only occur if the distribution of distances between plates is rather narrow, which is evidently not the case in our experiment. Therefore we tried a number of distributions of $d$ such as Gaussian and log-Gaussian ones to fit the experimental DDs of Fig. 3B without finding any satisfactory matching. The best fits were obtained with an empirically chosen discrete distribution of $d$ (Fig. 4). These best fits of calculated DDs to the experimental ones are shown in Fig.3B by solid lines and they describe the experimental DDs very well. The mean distance between planes in these simulations is 4.1 µm which agrees rather well with direct measurements ( ~3.8 µm and ~4.5 µm). In summary, the data in Fig.3B are well explained by simple mechanical hindrance exerted by the glass plates on the diffusive displacement in the IL film.

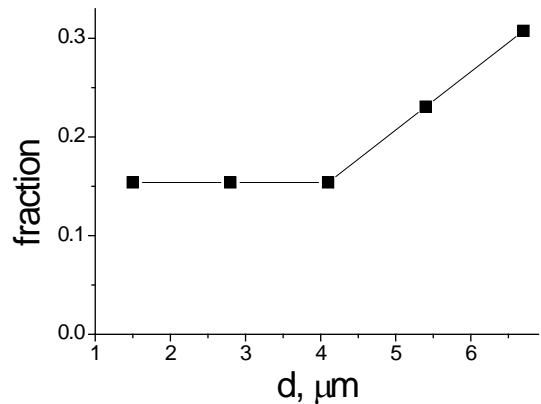

**Figure 4.** Distance size distribution used to fit DDs for the diffusion normal to the plates (Fig. 3B).

The DDs in Fig. 3A that arise from diffusive displacements in the IL films along the plates exhibit no dependence on the diffusion time and are thereby not influenced by any restrictions. This feature is as expected for a free diffusion of ions along the IL films. Yet, in remarkable contrast, diffusion along the plates seems to be characterized by a diffusion coefficient $D_{\parallel} = D^*$ that is by a factor of ~1.86 *higher* than the value of $D_0$ observed in the bulk! This increase is so large that it is comparable to that obtained in the bulk by increasing the temperature by *ca* 15 K.

Temperature dependence of $D^*$ shown in Fig. 5. It is seen that $D^*$ is higher than $D_0$ in the whole temperature interval by the same factor. The dependence has the same slope as $D_0(T)$, therefore it characterizes with the same activation energy for diffusion as EA in bulk ($E_D = 32.9$ kJ/mol).

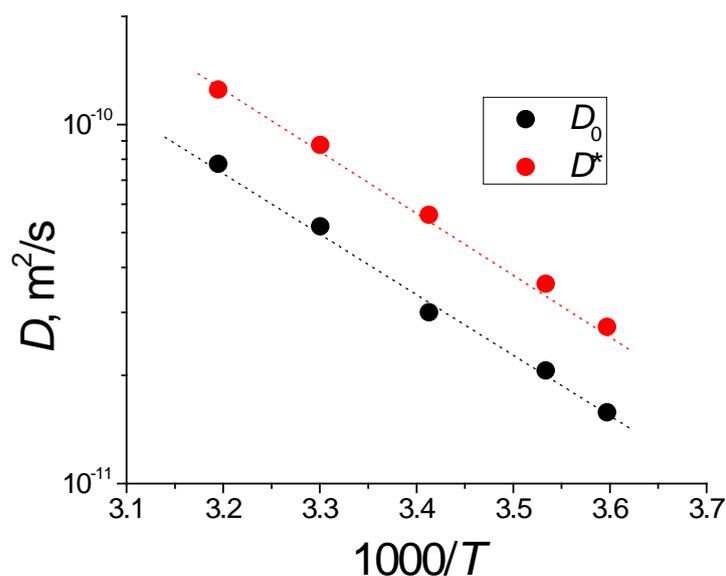

**Figure 5.** Arrhenius plots of $D^*$ and $D_0$.

$^1$H NMR $T_1$ in bulk demonstrated exponential forms of decays and showed that $T_1$ of protons of different chemical groups close to that in the confinement (Table 1). Transverse NMR relaxation of protons of -$CH_2$-, -$CH_3$ and -$NH_3^+$ groups demonstrated also exponential decays. After confinement the transverse relaxation decays were exponential for protons of -$NH_3^+$ groups, while non-exponential for protons of -$CH_2$- and -$CH_3$ groups. $T_2$ values for bulk and confined EAN also presented in Table 1. $T_2$ of different groups are decreased by factors of ~8.8 (-$CH_2$-), ~12.1 (-$CH_3$) and ~22 (-$NH_3^+$). There are a number of mechanisms leading to shortening of $T_2$ relaxation time.[40,41] However, in our case the maximum effect is offered to the exchangeable protons of -$NH_3^+$, which experiences chemical exchange between magnetically non-equivalent sites with following equilibrium:[42]

$$CH_3CH_2NH_3^+ + NO_3 \Leftrightarrow CH_3CH_2NH_2 + HNO_3.$$

Major decrease of $T_2$ (-$NH_3^+$) shows that confinement of EAN between polar glass-plates leads to accelerated exchange of protons of –$NH_3$ groups.

After having made this remarkable observation, it felt imperative to consider and rule out possible experimental artifacts. Hence, the experimental setups and samples were carefully tested. Particular attention was paid to the possible effect of water; since the stack contained many glass plates, one could assume that water adsorbed at the glass surface may have dissolved in the IL films and contributed to faster molecular dynamics. To exclude this as a significant contribution, the diffusion coefficient in bulk EAN with increasing water content has been measured.

**Table.** $^1$H NMR $T_1$ and $T_2$ relaxation times (s) of protons of different chemical groups of ethylammonium cation at 293 K, measured for bulk EAN and for EAN confined between the glass plates.

| EAN cation groups | -CH$_2$- | -CH$_3$ | -NH$_3^+$ |
|---|---|---|---|
| **Bulk $T_1$** | 0.67 ± 0.01 | 0.88 ± 0.01 | 0.39 ± 0.01 |
| **Confined $T_1$** | 0.79 ± 0.01 | 0.97 ± 0.01 | 0.49 ± 0.01 |
| **Bulk $T_2$** | 0.31 ± 0.01 | 0.51 ± 0.01 | 0.110 ± 0.05 |
| **Confined $T_2$ (averaged)** | 0.050 ± 0.001 | 0.079 ± 0.001 | 0.005 ± 0.0005 |

From those data and from the $^1$H NMR spectra, from which the approximate water content in EAN could be estimated, we can conclude that the effect of water to the observed phenomenon is negligible (approximately 8% water would be required to modify the diffusivity to the required extent, whereas the water content was shown to be <<0.1%). In summary, we conclude that the observed acceleration of diffusion along the confining glass plates is not an artifact, but an utterly surprising molecular effect. Below, we provide, with the support of some additional data, a tentative explanation to this observation.

Previous studies have established the notion that some surfaces can exert an effect on ILs that permeates the bulk for tens of nanometers. This effect was typically described as an imposed molecular order.[9,11-15] Other studies have shown that changes in short-range molecular order can have an influence on the macroscopic (displacements over micrometers) self-diffusion of the cations and anions in ILs.[22] On this basis, we provide a tentative interpretation of our findings.

The diffusion path length over 3 ms with the $D_0 = 3 \cdot 10^{-11}$ m$^2$/s, bulk diffusion coefficient, is $(2 \cdot D \cdot t_d)^{0.5}$ ~ 0.42 μm, or, if we take $D^* = 5.6 \cdot 10^{-11}$ m$^2$/s, the diffusion path is ~ 0.58 μm. However, DDs obtained for diffusion along the plates at 3 ms are single-exponential and coincide with the experimental point obtained at longer diffusion times (Fig.3A). That means that EAN diffusion is the same in the center of layer and near the plates (in proximity of ~0.5 μm to plates). Therefore, diffusion is accelerated not only in thin layers near surfaces, but in the whole thickness of the layer, even at the distance as far as ~2 μm from the surface.

Regarding the molecular order within the surface layer, if that is orientational akin to that exhibited by liquid crystals it may leave an imprint in from of NMR line splitting. This is the case if the structure within the layer is similar to that in liquid crystal phases with non-zero second-rank order parameter $S$, such as lamellar/smectic phases. In contrast, phases with $S = 0$, such as cubic LCs, do not exhibit line splitting. To investigate this, we recorded $^2$H NMR spectra of Sample 3 prepared with EAN-D$_5$. There is no detectable quadrupolar splitting in the $^2$H NMR spectra. This finding suggests that either the surface layer is thin, in which case the $^2$H NMR signal from it remains hidden in the background (at λ = 60 nm, is only ~ 3% of the NMR signal is arising the surface layer), or that the molecular order is such that the order parameter $S$ remains zero.

The diffusivity of EA cation is clearly shown to be isotropic when confined between glass plates. The bulk diffusion properties are highly consistent with the literature consensus of a sponge-like self-assembly structure. Consideration of spin echo data, where no significant splitting is observed further allows the conclusion to be drawn that the majority of cations are unaffected by the glass surfaces. Water is often present in ionic liquids and is fully compatible with EAN. It is categorically shown by exhaustive measurements that water associated with the polar glass plates cannot be solely responsible for the dramatic change in diffusivity compared to the bulk. The same time, the confinement leads to acceleration in exchange of -NH$_3$ protons in the whole layer. The only remaining possible explanation for the accelerated diffusion is that the structure or structural parameters of whole confined EAN is different in comparison with bulk EAN. For the sponge-like structure, which often attributed to the bulk phase of EAN[9-11], is characteristic hindered diffusion because of curvilinear trajectories of ions and small-size of connecting channels, while it is isotropic in the micrometer-scale range. At the same time sponge-like phase may change their structural parameters, while larger channels might allow an enhanced mobility.[43] Another known isotropic stricture, micellar, suggested formation of large aggregates of ions that will also hinder diffusion, but if size of micelle is not large, it can demonstrate accelerated diffusion and proton exchange for the –NH$_3$ groups. However, micellar structures of EAN have been observed only in EAN mixtures.[8] Furthermore, change in the phase of EAN usually leads to the change of activation energy for diffusion,[22] while in our case, $E_D$ is the same for $D_0$ and $D^*$. This confirm that the sponge-like phase of bulk EAN is saved in the confinement, but their effective "channels" increases that provides enhancement for diffusive transport and for accelerated –NH$_3$ protons exchange. Having made this experimental observation, further theoretical and experimental studies are required, for example on the effect of roughness, surface polarity and changing the ionic liquid constituents. Given that the liquid properties are rather altered, these results have strong implications for interface intensive applications of ILs, such as lubricants and in electrochemical systems.

**Acknowledgements**

The Knut and Alice Wallenberg foundation (project number KAW 2012.0078) and the Swedish Research Council (project numbers 621-2013-5171 (OA), 621-2011-4600 and 621-2014-4694 (SG), 621-2011-4361 (MR)) are gratefully acknowledged for their financial support. The Foundation in memory of J. C. and Seth M. Kempe and the laboratory fond at LTU are gratefully acknowledged for providing grants, from which a Bruker Aeon/Avance III NMR spectrometer at LTU has been purchased. NMR measurements were also partly carried out on the Bruker Avance III NMR spectrometer of the Federal Centre of Collective Facilities of Kazan Federal University, Russia.